\documentclass[pra,showpacs,superscriptaddress]{revtex4}
\usepackage{graphicx}% Include figure files
\usepackage{bm}

\begin{document}

\title{Counter-propagating entangled photons from a waveguide with periodic
nonlinearity}

\author{Mark~C.~Booth}
\email{mbooth@bu.edu} \homepage[Quantum Imaging Laboratory
homepage:~]{http://www.bu.edu/qil} \affiliation{Quantum Imaging
Laboratory, Department of Biomedical Engineering,}

\author{Mete~Atat\"{u}re}
\affiliation{Quantum Imaging Laboratory, Department of Physics,}

\author{Giovanni~Di~Giuseppe}
\altaffiliation[Also at~]{Istituto Elettrotecnico Nazionale {\it
G.~Ferraris}, Strada delle Cache 91, I-10153 Torino, Italy.}
\affiliation{Quantum Imaging Laboratory, Department of Electrical
\& Computer Engineering, Boston University, 8 Saint Mary's Street,
Boston, Massachusetts 02215}

\author{Alexander~V.~Sergienko}
\affiliation{Quantum Imaging Laboratory, Department of Physics,}
\affiliation{Quantum Imaging Laboratory, Department of Electrical
\& Computer Engineering, Boston University, 8 Saint Mary's Street,
Boston, Massachusetts 02215}

\author{Bahaa~E.~A.~Saleh}
\affiliation{Quantum Imaging Laboratory, Department of Electrical
\& Computer Engineering, Boston University, 8 Saint Mary's Street,
Boston, Massachusetts 02215}

\author{Malvin~C.~Teich}
\affiliation{Quantum Imaging Laboratory, Department of Biomedical
Engineering,} \affiliation{Quantum Imaging Laboratory, Department
of Physics,} \affiliation{Quantum Imaging Laboratory, Department
of Electrical \& Computer Engineering, Boston University, 8 Saint
Mary's Street, Boston, Massachusetts 02215}

\date{\today}

\renewcommand{\baselinestretch}{2}\small\normalsize

\begin{abstract}
The conditions required for spontaneous parametric down-conversion
in a waveguide with periodic nonlinearity in the presence of an
unguided pump field are established. Control of the periodic
nonlinearity and the physical properties of the waveguide permits
the quasi-phase matching equations that describe
counter-propagating guided signal and idler beams to be satisfied.
We compare the tuning curves and spectral properties of such
counter-propagating beams to those for co-propagating beams under
typical experimental conditions. We find that the
counter-propagating beams exhibit narrow bandwidth permitting the
generation of quantum states that possess discrete-frequency
entanglement. Such states may be useful for experiments in quantum
optics and technologies that benefit from frequency entanglement.
\end{abstract}


\maketitle
%\newpage
\section{Introduction\label{sec.intro}}
To achieve successful frequency mixing of optical waves in a
nonlinear medium, it is essential that the relative phase between
the interacting waves vanish. In the bulk nonlinear media most
often encountered in nonlinear optics, {\em birefringent}
phase-matching is used to correct for the phase
difference~\cite{Giordmaine62,Maker62}. The difference in
refractive index for each polarization of the interacting waves
serves to correct the phase mismatch and ensure effective coupling
among the waves. However, this method fails when the medium does
not exhibit sufficient birefringence at the interacting
wavelengths. As an alternative to birefringent phase-matching,
waveguide structures can also be utilized to achieve wave
coupling~\cite{Shen_nonlinearoptics}. Coupled-mode theory
describes the power transfer among guided waves and relies on
matching the propagation constants of the interacting fields along
the direction of travel, rather than the wavevector for each
optical wave. By careful selection of the waveguide dimensions and
other properties such as the index profile, the modal structure of
the interacting waves can be altered such that efficient frequency
mixing can be realized within the waveguide.

Yet another method to correct for phase mismatch is to introduce
periodic structure into the nonlinear medium, which results in
what is termed {\em quasi-phase matching} (QPM).  This approach
was independently proposed by Armstrong {\em et
al.}~\cite{Armstrong62} and Franken and Ward~\cite{Franken63} in
the early 1960's. Of particular interest is a technique for
modulating the sign of the second-order nonlinearity throughout
the medium to produce a so-called {\em periodically poled
material}. Since any interaction within the transparency region of
such a periodically poled material can be non-critically phase
matched at a specific temperature via QPM, it is possible to
utilize the largest value of the effective nonlinear coefficient
for a given material and thus increase the overall conversion
efficiency. In the family of commonly used nonlinear media,
lithium niobate (${\rm LiNbO_3}$) boasts one of the largest
nonlinear coefficients $d_{33}$ for all beams having extraordinary
polarization; it is approximately 6 times greater than the
$d_{31}$ coefficient used in birefringent
phase-matching~\cite{Baldi95}. The result is a theoretical
enhancement of $(2d_{33}/\pi d_{31})^2 \sim 20$ for QPM over
birefringent phase-matching~\cite{Myers95}. Many of the advantages
of QPM in lithium niobate have been outlined
elsewhere~\cite{Fejer92} and results show that periodically poled
lithium niobate (PPLN) is a natural choice for many experiments.

The enhancement in nonlinear-interaction efficiency promised by
the use of PPLN has raised interest in the quantum-optics
community where there has been a growing interest in new
ultrabright sources of entangled photons~\cite{Tanzilli01_EL}.
Entangled photons, which may be generated through the process of
spontaneous parametric down-conversion (SPDC) in a crystal with
$\chi^{(2)}$ nonlinearity, has long been in the spotlight for
quantum-optics experiments~\cite{Perina94,Mandel95}. SPDC,
however, suffers from low conversion efficiency, on the order of
$10^{-9}$ entangled-photon pairs per mode per pump photon, which
ultimately limits their use in many practical applications.
Lithium niobate offers the promise of increased photon-pair
production and, with the integration of a waveguide structure,
control of the spatial characteristics of the down-converted
photons while still maintaining a substantial increase in
conversion efficiency~\cite{Tanzilli01_EL,Banaszek01}.

It turns out, however, that the use of a waveguide structure, in
conjunction with periodic poling, offers yet another critically
important feature: the possibility of generating
counter-propagating signal and idler photons. Counter-propagation
has been previously considered in the context of surface
birefringence in nanostructure semiconductors~\cite{DeRossi02}.

In this paper, we consider the conditions required for generating
counter-propagating photon beams by quasi-phase matching in
waveguides with periodic nonlinearity and we explore the unique
properties of this source of light.  Such a source of entangled
photons would be immediately useful in quantum-interference
experiments that make use of spatial filters~\cite{Atature01_PRA}.
Such filters are often used to restore visibility in these kinds
of experiments but they carry the price of a significant reduction
in the photon-pair rate.

\section{SPDC in a single-mode waveguide\label{sec.single_mode}}
By virtue of a relatively weak interaction, time-dependent
perturbation theory leads to a two-photon state generated by SPDC
at the crystal output that is given by~\cite{Klyshko80}
\begin{equation}
  \label{eq:Psi-general}
  |\Psi^{(2)}\rangle \sim \int_V {\rm d}V \int {\rm d}t\,\, \chi^{(2)}\!(z)\,
  \hat{E}_{\rm p}^{(+)}\!({\bf r},t)\, \hat{E}_{\rm s}^{(-)}\!({\bf r},t)\,
  \hat{E}_{\rm i}^{(-)}\!({\bf r},t)\,|0\rangle,
\end{equation}
where $V$ is the interaction volume, $\chi^{(2)}$ is the
second-order susceptibility of the medium, ${\bf r}$ is the
position vector, and $\hat{E}_j^{(\pm)}$ is the positive- or
negative-frequency part of the pump, signal, or idler
electric-field operator ($j=$ p, s, i), respectively. In a
waveguide, the single-mode signal and idler fields can be
described by
\begin{equation}
  \label{eq:signal-idler}
  \hat{E}_j^{(-)}\!({\bf r},t) = \int {\rm d}\omega_j \,
  u_0({\bf x};\omega_j) \, e^{-{\rm i}\beta_j z} \, e^{{\rm i}\omega_j t} \,
  \hat{a}^\dagger_j(\omega_j,\beta_j),
\end{equation}
where $u_0$ corresponds to the fundamental mode of the waveguide,
${\bf x}$ is the transverse position vector, $\beta_j =
\beta_j(\omega)$ are the propagation constants, and
$\hat{a}^\dagger_j$ are the creation operators, with $j=$ s, i,
respectively. As long as the propagation constants for signal and
idler fields, $\beta_{\rm s}$ and $\beta_{\rm i}$, are larger than
a critical propagation constant $\beta_{\rm c}$, determined by the
waveguide properties, there is only one fundamental mode available
for the propagation of each of the signal and idler fields.  If
such a condition is met for both the signal and idler fields, then
this corresponds to a single-mode waveguide for these fields, as
illustrated in Fig.~\ref{fig:schematic} for an arbitrary pump
field. This assumption is retained throughout the remainder of
this paper.

If it is assumed, furthermore, that the complex pump field is
classical:
\begin{equation}
    \label{eq:pump-field-general}
    E_{\rm p}({\bf r},t) = \int {\rm d}\omega_p \, \mathcal{E_{\rm p}}({\bf x};\omega_{\rm p}) \,
    e^{{\rm i} \beta_{\rm p} z} \, e^{-{\rm i} \omega_{\rm p} t}\,,
\end{equation}
where $\mathcal{E_{\rm p}}$ is the transverse and spectral profile
of the pump wave, Eqs.~(\ref{eq:Psi-general}) -
(\ref{eq:pump-field-general}) then lead to a two-photon quantum
state given by
\begin{equation}
  \label{eq:Psi-specific}
  |\Psi^{(2)}\rangle \sim \int_{\rm A} {\rm d}{\bf x} \, {\rm d}z \int {\rm d}t
  \int {\rm d}\omega_{\rm p} \, {\rm d}\omega_{\rm s} \, {\rm d}\omega_{\rm i}
  \,\, \mathcal{E_{\rm p}}({\bf x};\omega_{\rm p}) \, u_0({\bf x};\omega_{\rm s}) \,
  u_0({\bf x};\omega_{\rm i}) \,
  \chi^{(2)}\!(z) \, e^{{\rm i} \Delta \beta z} \, e^{-{\rm i}(\omega_{\rm p} - \omega_{\rm s} - \omega_{\rm i})t} \,
  \hat{a}_{\rm s}^\dagger \hat{a}_{\rm i}^\dagger\,|0\rangle,
\end{equation}
where $A$ is the transverse plane spanned by the vector ${\bf x}$
and where
\begin{equation}
    \label{eq:delta_beta}
        \Delta \beta = \beta_{\rm p} - \beta_{\rm s} - \beta_{\rm i}
\end{equation}
is the phase mismatch between the three fields that must vanish
for perfect phase-matching.

We can further simplify Eq.~(\ref{eq:Psi-specific}) by carrying
out the integration over $t$. This yields the condition
\begin{equation}
  \label{eq:energy-conservation}
    \omega_{\rm p} - \omega_{\rm s} - \omega_{\rm i} = 0
\end{equation}
and, using the Fourier-transform definition
\begin{equation}
  \label{eq:FT-chi}
    \tilde{\chi}^{(2)}\!(\Delta \beta) = \int {\rm d}z \, e^{{\rm i} z \Delta \beta}\,
    \chi^{(2)}\!(z),
\end{equation}
we obtain
\begin{equation}
  \label{eq:Psi-specific-final}
  |\Psi^{(2)}\rangle \sim \int {\rm d}{\bf x} \int {\rm d}\omega_{\rm s} \, {\rm d}\omega_{\rm i}
  \,\, \mathcal{E_{\rm p}}({\bf x};\omega_{\rm s} + \omega_{\rm i}) \, u_0({\bf x};\omega_{\rm s}) \,
  u_0({\bf x};\omega_{\rm i}) \,
  \tilde{\chi}^{(2)}\!(\Delta \beta) \, \hat{a}_{\rm s}^\dagger
  \hat{a}_{\rm i}^\dagger\,|0\rangle
\end{equation}
for the two-photon state describing down-converted light in a
single-mode waveguide structure.

Finally, applying the creation operators to the vacuum yields
\begin{equation}
  \label{eq:Psi-2pump}
  |\Psi^{(2)}\rangle \sim \int{\rm d}\omega_{\rm s} \, {\rm d}\omega_{\rm i}
  \, \Phi(\omega_{\rm s},\omega_{\rm i}) \,
  |1,1\rangle,
\end{equation}
with a state function $\Phi$~\cite{Giuseppe02} given by
\begin{equation}
  \label{eq:SPDC-BW}
    \Phi(\omega_{\rm s},\omega_{\rm i}) = \int_{\rm A}{\rm d}{\bf x} \,
  \mathcal{E}_{\rm p}({\bf x};\omega_{\rm s} + \omega_{\rm i})\, u_0({\bf x};\omega_{\rm s}) \,
  u_0({\bf x};\omega_{\rm i})\, \tilde{\chi}^{(2)}\!(\Delta
  \beta).
\end{equation}
The ket $|1,1\rangle$ represents a single photon in the signal
mode and a single photon in the idler mode. Since the
down-converted light is guided in only a single spatial mode, the
magnitude-squared of the state function, $|\Phi(\omega_{\rm
s},\omega_{\rm i})|^2$, represents the spectrum of the SPDC, which
will be considered subsequently.

The theory presented to this point is general and can be applied
to any structure with an arbitrary nonlinearity profile. In this
paper, we focus on a second-order susceptibility that is
periodically modulated in the $z$-direction, so that
$\chi^{(2)}\!(z)$ can be written as a Fourier series and takes the
form
\begin{equation}
  \label{eq:X-modulation}
  \chi^{(2)}\!(z) = \chi_0^{(2)} \sum_m G_m e^{{\rm i} K_m z}.
\end{equation}
Here $\emph m$ corresponds to the $m^{\rm th}$ component of the
Fourier expansion with coefficient $G_m$ and wavenumber $K_m =
2\pi m / \Lambda$, where $\Lambda$ is the periodicity of the
modulation.  Since the phase mismatch $\Delta \beta$ among the
propagation constants of the three fields is most often not zero,
we must use the contribution $K_m$, associated with the modulation
of the nonlinearity in the material, to achieve perfect
quasi-phase matching in the waveguide [$\Delta \beta = K_m$ in
Eq.~(\ref{eq:delta_beta})].  We may therefore define a phase
mismatch $\Delta \beta'$ in the presence of the grating vector
$K_m$ given by
\begin{equation}
  \label{eq:quasi-phase-matching}
  \Delta \beta' = \beta_{\rm p} - \beta_{\rm s} - \beta_{\rm i} - K_m.
\end{equation}
When $\Delta \beta' = 0$, this is referred to as perfect {\em
quasi}-phase matching.

For a nonlinearity with a single periodicity, $K_{\pm m} = \pm
K_m$ so that negative values of $m$ may also yield meaningful
solutions to the QPM conditions given in
Eq.~(\ref{eq:quasi-phase-matching}). In conjunction with the
multiple signal-idler wavelength combinations generated with SPDC,
contributions from both positive and negative values of $m$ can be
simultaneously realized.  This is in contradistinction to
three-wave mixing problems such as second harmonic
generation~\cite{Fejer92}, where the two input fields are fixed
and yield only one output field. In this latter case, the grating
vector corresponding to the negative value of $m$ is not utilized
since it does not yield a simultaneous phase-matching solution.

\section{Counter-propagating SPDC in a PPLN waveguide\label{sec.counter_prop}}
We continue our analysis by considering the specific example of a
PPLN waveguide and proceed to describe some interesting properties
of single-mode, counter-propagating signal and idler beams, under
various conditions.

Without loss of generality, we consider a waveguide in two
dimensions~\cite{Saleh91_waveguides}: $x$ as the transverse
(guiding) dimension and $z$ as the longitudinal (propagating)
dimension. We further assume that the pump field in
Eq.~({\ref{eq:pump-field-general}) is a monochromatic unguided and
undepleted plane wave with a transverse spatial profile given by:
\begin{equation}
  \label{eq:pump-field}
    \mathcal{E_{\rm p}}(x;\omega_{\rm p})= A_{\rm p}\cos[(k_{\rm p}\sin\theta) x ],
\end{equation}
where $A_{\rm p}$ is a constant and $k_{\rm p}$ is the pump
wavenumber. Such a field can be realized by illuminating the
waveguide with two identical coherent fields at angles $\pm
\theta$ from the $z$-axis. An illustration of a planar dielectric
waveguide structure pumped in this geometry is shown in
Fig.~\ref{fig:unguided-pump}.  Since the power of the pump beam
can be increased to any desired level, loss of pump beam energy is
not critical for effective operation.  Our example of an unguided
pump beam is distinct from a scheme for co-propagating difference
frequency generation, in which both the pump beam and signal
photon are guided and a co-propagating idler photon is allowed to
radiate into the substrate~\cite{Baldi95}.

To compute tuning curves from Eq.~(\ref{eq:energy-conservation})
and Eq.~(\ref{eq:quasi-phase-matching}), where the pump beam is
incident on the waveguide structure at an angle $\theta$, we use
$\beta_{\rm p} = k_{\rm p} \cos\theta$. We allow the propagation
constant $\beta_i$ for the idler field to take on a negative value
to permit counter-propagation. Finally, to compute the specific
values of $\beta_j$ for each field, we use the set of Sellmeier
equations for lithium niobate~\cite{Crystals.Handbook} to
determine the refractive index for extra-ordinarily polarized (e)
waves in the e~$\rightarrow$~e~+~e interaction under
consideration. Although Eq.~(\ref{eq:quasi-phase-matching}) only
describes perfect phase-matching ($\Delta \beta' = 0$), we are
nonetheless able to determine many interesting properties of
counter-propagating signal and idler photons.  These are presented
as tuning curves, representing various properties of the
down-converted light as a function of the pump-beam incidence
angle.

To investigate the relationship of the poling period and the
pump-beam incidence angle for quasi-phase matching, we first
consider only the degenerate case where the pump field gives rise
to counter-propagating signal and idler photons at twice the pump
wavelength ($\lambda_{\rm s} = \lambda_{\rm i} = 2\lambda_{\rm p}
=$ 1064 nm). Figure~\ref{fig:degen} shows the required poling
period $\Lambda$ that allows for perfect QPM at any pump-beam
incidence angle. Immediately evident from this plot is that for a
pump beam that is coupled into the PPLN waveguide at $65^\circ$,
one would require a poling period on the order of the pump
wavelength (0.5 $\mu$m). This observation remains true for
pump-beam incidence angles less than $65^\circ$, including the
case for normal incidence at $0^\circ$. Since the smallest poling
period experimentally achievable with lithium niobate is currently
$\sim$4~$\mu\rm{m}$~\cite{Batchko99}, this configuration is not
currently realizable.

To experimentally realize phase-matching for counter-propagating
fields in an optical parametric oscillator or amplifier (OPO/OPA)
at angles near normal incidence, researchers have been forced to
utilize materials such as gallium arsenide, in which it is
possible to construct sub-micron multilayer or asymmetric
quantum-well domain structures using semiconductor growth
techniques. The second-order susceptibility can then be spatially
modulated between contiguous layers at the dimension required to
successfully achieve phase-matching. By varying the pump beam
angle in such devices, QPM may be realized over large tuning
ranges in OPOs and OPAs~\cite{Ding95}.

The use of an unguided pump avoids the necessity for sub-micron
poling periods to achieve counter-propagation in PPLN waveguides.
To illustrate this, consider the particular first-order poling
period $\Lambda = 6.8~\mu\rm{m}$ that corresponds to second
harmonic generation at 532~nm and, therefore, degenerate
down-conversion to 1064~nm. For this poling period, we find from
Fig.~\ref{fig:degen} that if we pump the waveguide at an angle of
$\sim$88.2$^\circ$, we can generate counter-propagating signal and
idler beams. When the pump beam is at $90^\circ$, $\Lambda
\rightarrow \infty$ ($K_m \rightarrow 0$) so that QPM is no longer
required for down-conversion and a birefringent interaction will
result in counter-propagating signal and idler
photons~\cite{Fejer92}.

Most interesting is non-degenerate down-conversion where we
select, as a particular example, the signal-idler combination of
810 nm and 1550 nm, respectively. From the tuning curves in
Fig.~\ref{fig:nondegen} we find that for the same poling period
considered in Fig.~\ref{fig:degen}, $\Lambda = 6.8~\mu\rm{m}$, one
can satisfy QPM at two pump angles, namely $\sim$70.4$^\circ$ and
$\sim$74.6$^\circ$, as a result of the dual-directionality of the
grating vector $K_m$. Since there are two angles that satisfy QPM
for non-degenerate down-conversion, we expect that for a single
angle, the dual-directionality of the grating vector would allow
for different non-degenerate wavelength combinations. In
Fig.~\ref{fig:lambda_v_angle}(a), all possible signal-idler
wavelength combinations are plotted as the pump-beam incidence
angle is varied from $65^\circ$ to $90^\circ$. For this plot, the
poling period is fixed at $\Lambda = 6.8~\mu\rm{m}$ and the
corresponding tuning curves are plotted for grating vector values
with order $m =$ 0, $\pm$1, $\pm$2, and $\pm$3.  The selection of
tuning curves for experimental implementation is dependent on the
nature of the periodicity of the nonlinearity and on the strength
of the coefficients $G_m$ in Eq.~(\ref{eq:X-modulation}).

As an example, a $\chi^{(2)}$ that is modulated as a square wave
in the $z$-direction, as illustrated in
Fig.~\ref{fig:unguided-pump}, has dominant Fourier components of
order $\pm$1, for which tuning curves are shown in
Fig.~\ref{fig:lambda_v_angle}(b). If we consider an example where
the pump beam angle is $80^\circ$, there are signal-idler
wavelength combinations of approximately 880 nm/1350 nm ($m=$~-1,
dashed curve, open circles) and 930 nm/1240 nm ($m=$~+1, solid
curve, solid circles). This quantum state can be represented by
\begin{equation}
  \label{eq:lambda_superposition}
  |\Psi^{(2)}\rangle \sim c_1|880,1350\rangle +
                          c_2|930,1240\rangle,
\end{equation}
where the constants $c_1$ and $c_2$ are determined mainly by the
pump properties. By simply selecting the pump-beam incidence angle
to be $74.6^\circ$, for example, the two-photon quantum state
given above can be readily tuned to new signal-idler wavelength
combinations of 810 nm/1550 nm and 860 nm/1380 nm.

The generation of a superposition of two counter-propagating
non-degenerate photons pairs occurs naturally within the PPLN
waveguide structure and this is a novel property as it stands.
Beyond changing the pump-beam incidence angle, control over the
down-converted light is further afforded by adjusting the poling
period. As the poling period is reduced, the difference between
the two signal wavelengths and the two idler wavelengths turns out
to increase, i.e., the curves separate from one another. As one
increases the poling period or chooses the longer third-order
poling period, the signal photons and idler photons become more
similar in wavelength, i.e., the curves approach each other.

\section{SPDC Spectrum in a PPLN waveguide\label{sec.bandwidth}}
The preceding section considered only the idealized phase-matching
conditions provided by Eq.~(\ref{eq:quasi-phase-matching}). In
this section, we consider effects on the spatio-temporal
distribution of down-converted light imparted by the finite
crystal length and the modal structure of the waveguide. The
spectrum of down-conversion at the output of the waveguide is
computed from the magnitude-squared of the state function given in
Eq.~(\ref{eq:SPDC-BW}).

As a specific example, and for convenience of calculation, we
consider the TE modes of a planar graded-index (GRIN) slab
waveguide as illustrated in Fig.~\ref{fig:unguided-pump}, with a
parabolic index profile~\cite{Saleh91_waveguides} along the
transverse $x$-axis given by
\begin{equation}
  \label{eq:parabolic-index-profile}
  n^2(x) = n_0^2\left[1-\alpha^2 x^2\right],
\end{equation}
where $n_0$ is the maximum core index and $n_0 \alpha^2$ is the
rate of change of refractive index with position, ${\rm
d}n(x)/{\rm d}x$. For this case, the fundamental transverse mode
becomes~\cite{Ghatak98}
\begin{equation}
  \label{eq:phi}
    u_0(x;\omega_j) = C \, {\rm exp}\!\!\left[-\frac{(\gamma_j x)^2}{2}\right],
\end{equation}
where $C$ is a constant and where
\begin{equation}
    \label{eq:transverse-mode-constants}
    \gamma_j = \left[ \frac{n_0 \omega_j}{c} \, \alpha
    \right]^{\frac{1}{2}}, \quad\quad
    \beta_j =  \frac{n_0 \omega_j}{c} \left[ 1 - \frac{c}{n_0 \omega_j } \,\alpha\right]^{\frac{1}{2}}.
\end{equation}
Single-mode operation is achieved when
\begin{equation}
    \label{eq:beta-critical}
    \beta_{{\rm s,i}} > \beta_{\rm c} = \frac{n_0 \omega_{\rm s,i}}{c}
    \left[ 1 - \frac{3 c}{n_0 \omega_j }\, \alpha\right]^{\frac{1}{2}}
\end{equation}
and we assume that the down-converted light couples only to the
fundamental mode of the parabolic waveguide.

By substituting into Eq.~(\ref{eq:SPDC-BW}) the transverse mode
profile given by Eq.~(\ref{eq:phi}), the unguided pump field given
by Eq.~(\ref{eq:pump-field}), and the first-order periodic
nonlinearity found from Eq.~(\ref{eq:X-modulation}), we find the
following expression for the SPDC spectrum:
\begin{equation}
  \label{eq:phi-dual-pump-pp-material}
  |\Phi(\omega_{\rm s},\omega_{\rm i})|^2 = A \, \sqrt{\frac{\pi}{\gamma_{\rm s}^2 + \gamma_{\rm i}^2}}
  \,\,\, {\rm exp}\!\!\left[-\frac{\zeta^2}{4 \gamma_{{\rm
  eff}}^2}\right]
   \left\{ \,
   {\rm sinc} \left[ \left( \Delta \beta + \frac{2 \pi}{\Lambda} \right) \frac{L}{2} \right]
  +
  {\rm sinc} \left[ \left( \Delta \beta - \frac{2 \pi}{\Lambda} \right) \frac{L}{2} \right]
  \, \right\},
\end{equation}
where $A$ is a constant, $\zeta = [n_0(\omega_{\rm s} +
\omega_{\rm i})/c] \sin\theta$, $\gamma_{{\rm eff}}^2 =
(1/2)(\gamma_{\rm s}^2 + \gamma_{\rm i}^2)$, and $L$ is the
crystal length. Effectively, the SPDC bandwidth is modulated by
the prefactor $[\pi/(\gamma_{\rm s}^2 + \gamma_{\rm
i}^2)]^{1/2}\,{\rm exp}(-\zeta^2/4 \gamma_{{\rm eff}}^2)$
associated with the waveguide and will therefore differ from that
for bulk media. The spectral profile of the photon pairs can
therefore be altered by the waveguide parameters. For the
waveguide parameters in the parabolic index-profile example
presented here, however, the value of this pre-factor remains
nearly constant over the computed bandwidth, so that waveguide
effects can be ignored, at least for the fundamental mode. If the
waveguide operating conditions admit multiple modes, however, the
SPDC bandwidth may be more sensitive to the waveguide properties.

We now consider the SPDC bandwidth, measured as the
full-width-half-maximum (FWHM) of the SPDC spectrum, for several
interesting configurations in a PPLN waveguide.
Figure~\ref{fig:SPDC_degen} presents a set of plots that show the
SPDC spectrum for (a) degenerate, co-propagating ($\theta =
0^\circ$), (b) degenerate, counter-propagating ($\theta =
88.2^\circ$), (c) non-degenerate, co-propagating ($\Lambda =
7.4~\mu{\rm m}$, $\theta = 0^\circ$), and (d) non-degenerate,
counter-propagating ($\theta = 69.7^\circ$) down-converted
photons.  In all cases, the PPLN waveguide is taken to have an
interaction length $L =$~1 mm, a poling period of 6.8~$\mu$m
(unless otherwise specified), and the pump wavelength is again
taken to be 532~nm. It is immediately obvious that the bandwidths
for the counter-propagating cases are narrower than those for the
co-propagating cases [compare Figs.~\ref{fig:SPDC_degen}(b) and
\ref{fig:SPDC_degen}(d) with Figs.~\ref{fig:SPDC_degen}(a) and
\ref{fig:SPDC_degen}(c)]. This narrow bandwidth arises since the
sum, rather than the difference, of the propagation constants
appears in the expression for $\Delta \beta'$~\cite{Fejer92}. For
degenerate signal and idler photons, the reduction in bandwidth
from 130~nm for the co-propagating case to 0.23~nm in the
counter-propagating case is almost three orders-of-magnitude. This
is far in excess of the one order-of-magnitude reduction in
bandwidth available via surface birefringence in a semiconductor
waveguide~\cite{DeRossi02}. The pronounced bandwidth reduction in
our PPLN structure arises from the strong dispersion of lithium
niobate. The narrow bandwidth of the signal and idler supports the
approximate discrete-frequency representation provided in
Eq.~(\ref{eq:lambda_superposition}) and indicates that nearly
discrete down-conversion wavelengths can indeed be realized
experimentally without having to resort to the use of narrow
spectral filters.

To highlight the distinction between the co-propagating and
counter-propagating cases, we present a plot of the SPDC bandwidth
ratio versus normalized signal wavelength ($\lambda_{\rm
s}/2\lambda_{\rm p}$) in Fig.~\ref{fig:SPDC_nondegen}.  The SPDC
bandwidth ratio is obtained by dividing the FWHM bandwidth for
each central signal wavelength by the FWHM bandwidth at
$\lambda_{\rm s}$ = 880 nm, which is the shortest wavelength used
for this simulation. The PPLN waveguide is taken to have an
interaction length $L =$~1~mm and again $\lambda_{\rm p} =$
532~nm. This plot illustrates the relative increase in bandwidth
for co-propagating and counter-propagating photons as the
nonlinear interaction approaches degeneracy ($\lambda_{\rm
s}/2\lambda_{\rm p} =$~1.00). This trend is similar in
birefringent phase-matching and arises from a change in the
refractive index sum or difference for signal and idler photons.
The change, however, is much more pronounced for co-propagating
photons. Note that the bandwidth ratio (and the bandwidth) of the
counter-propagating photons is always less than that for
co-propagating photons and changes far less as the signal-idler
wavelength combination changes.

Just as we did for the analysis presented in
Sec.~\ref{sec.counter_prop}, we generate tuning curves that
illustrate all possible signal-idler combinations for various
pump-beam incidence angles. Figure~\ref{fig:2D_SPDC_spec} shows
the tuning curve obtained for co-propagating signal and idler
photons in a 1-mm PPLN waveguide when the spectral distribution of
the down-converted light is incorporated in the model. To generate
such a curve for co-propagating photons, it is necessary to
increase the poling period in the simulation to 7.4 $\mu$m so that
non-degenerate solutions can be found as the pump-beam incidence
angle is varied. The grating vector $K_m$ is taken to be positive
($m=$~1) to facilitate comparison with previous work in the field.
The plot reveals the large bandwidth near degeneracy (1064~nm) and
the spectral characteristics of the down-converted light for all
wavelength combinations.

Finally, in Fig.~\ref{fig:2D_counter_SPDC_spec}, we present a
tuning curve for counter-propagating signal and idler photons in a
1-mm PPLN waveguide with a poling period of 6.8 $\mu$m.  In
analogy with the analysis carried out in
Sec.~\ref{sec.counter_prop}, we consider both the positive and
negative spatial components of the grating vector ($m=\pm$1). The
resulting plot is similar to that presented in
Fig.~\ref{fig:lambda_v_angle}(b) but now incorporates the spectral
distribution of the down-converted light.  Since the bandwidth of
the down-converted light is narrow for all signal-idler wavelength
combinations, and does not change appreciably as the photons
approach degeneracy, the curves remain narrow for all pump-beam
incidence angles.

\section{SPDC in a cladding-pumped nonlinear fiber\label{sec.fibers}}
Co-propagating entangled photons generated by spontaneous
parametric down-conversion have previously been observed in
periodically poled silica fibers (PPSFs) by directly coupling of
the pump beam~\cite{Bonfrate99}. Although the effective
nonlinearity $d_{\rm eff}$ in silica is less than that for lithium
niobate, PPSFs offer a longer interaction length $L$, a higher
damage intensity threshold $I$, and a lower refractive index $n$
so that they achieve a comparable figure of merit $d^2_{\rm
eff}L^2I/n^3$.  The shortest poling period currently available in
a D-shaped silica fiber is $\Lambda = 25~\mu$m, which was used for
second-harmonic generation at 422~nm with an average conversion
efficiency of $\sim$0.22\%~\cite{Kazansky97}.  It is expected that
this efficiency would be able to be increased by a factor of 100
by improvements in poling techniques.

As an alternate to the periodically poled silica fiber and to the
PPLN waveguide example presented in
Secs.~\ref{sec.counter_prop}~and~\ref{sec.bandwidth}, we present a
novel configuration in which we propose the use of a periodically
poled nonlinear optical fiber to generate counter-propagating
signal and idler photons. In a scheme analogous to that of a fiber
laser, the pump beam (which is guided in the fiber cladding)
couples to the core medium as it propagates. Since the core medium
has a modulated nonlinearity, as depicted in
Fig.~\ref{fig:nonlinear-fiber}, counter-propagating entangled
photons can be generated.  With improvement in the conversion
efficiency, the use of cladding-pumped nonlinear fibers such as
described above should prove an effective source of
counter-propagating entangled photons for quantum-optics
experiments and technologies.

\section{Conclusion}
In Sec.~\ref{sec.single_mode}, we discussed the conditions for
achieving spontaneous parametric down-conversion in a waveguide
with periodic nonlinearity. We extended this discussion in
Sec.~\ref{sec.counter_prop} and found that an unguided pump beam
in a periodically poled lithium niobate waveguide could give rise
to counter-propagating signal and idler photons.  In
Sec.~\ref{sec.bandwidth}, we considered a particular waveguide
structure and computed the spectrum of the down-converted light
for several interesting and important cases. The theoretical
results indicate that counter-propagating beams have a narrower
bandwidth than co-propagating beams, which makes it possible to
generate a superposition of two or more counter-propagating
non-degenerate photon pairs.  The quantum state generated thereby
exhibits {\em discrete} frequency entanglement and can be readily
altered in several ways: by tuning the pump-beam incidence angle,
by appropriately changing the pump field profile using, e.g., a
superposition of pump angles, and by engineering the periodicity
of the nonlinearity. Such a quantum state cannot be generated in
bulk nonlinear media, nor in media with periodic structures, but
they are generated naturally in media with both periodic
nonlinearity and a waveguiding structure. Although we primarily
discussed theoretical results for a PPLN waveguide, the results
are general and will also apply, for example, to a periodically
poled cladding-pumped fiber with $\chi^{(2)}$ nonlinearity.  Many
other novel configurations for the generation of
counter-propagating signal and idler photons can be envisioned.

It may also be possible to admix other levels of subtlety into the
quantum states described in this paper.  By carefully selecting an
appropriate nonlinear medium and crystal cut axes, for example,
polarization entanglement can be incorporated into the
down-converted photons. It has been shown in potassium titanyl
phosphate (KTP) that it is possible to create type-II second
harmonic conversion by use of the $d_{24}$ nonlinear coefficient
for $z$-cut crystals with the beam propagation along the
$x$-axis~\cite{Wang99}. Further studies along these lines will
certainly reveal other such new and interesting materials,
implementations, and devices.

\begin{acknowledgments}
This work was supported by the National Science Foundation; the
Center for Subsurface Sensing and Imaging Systems (CenSSIS), an
NSF Engineering Research Center; the Defense Advanced Research
Projects Agency (DARPA); and the David and Lucile Packard
Foundation.
\end{acknowledgments}

\bibliographystyle{apsrev}
\bibliography{qil-bib-1}

\begin{thebibliography}{24}
\expandafter\ifx\csname natexlab\endcsname\relax\def\natexlab#1{#1}\fi
\expandafter\ifx\csname bibnamefont\endcsname\relax
  \def\bibnamefont#1{#1}\fi
\expandafter\ifx\csname bibfnamefont\endcsname\relax
  \def\bibfnamefont#1{#1}\fi
\expandafter\ifx\csname citenamefont\endcsname\relax
  \def\citenamefont#1{#1}\fi
\expandafter\ifx\csname url\endcsname\relax
  \def\url#1{\texttt{#1}}\fi
\expandafter\ifx\csname urlprefix\endcsname\relax\def\urlprefix{URL }\fi
\providecommand{\bibinfo}[2]{#2}
\providecommand{\eprint}[2][]{\url{#2}}

\bibitem[{\citenamefont{Giordmaine}(1962)}]{Giordmaine62}
\bibinfo{author}{\bibfnamefont{J.~A.} \bibnamefont{Giordmaine}},
  \bibinfo{journal}{Phys. Rev. Lett.} \textbf{\bibinfo{volume}{8}},
  \bibinfo{pages}{19} (\bibinfo{year}{1962}).

\bibitem[{\citenamefont{Maker et~al.}(1962)\citenamefont{Maker, Terhune,
  Nisenoff, and Savage}}]{Maker62}
\bibinfo{author}{\bibfnamefont{P.~D.} \bibnamefont{Maker}},
  \bibinfo{author}{\bibfnamefont{R.~W.} \bibnamefont{Terhune}},
  \bibinfo{author}{\bibfnamefont{M.}~\bibnamefont{Nisenoff}}, \bibnamefont{and}
  \bibinfo{author}{\bibfnamefont{C.~M.} \bibnamefont{Savage}},
  \bibinfo{journal}{Phys. Rev. Lett.} \textbf{\bibinfo{volume}{8}},
  \bibinfo{pages}{21} (\bibinfo{year}{1962}).

\bibitem[{\citenamefont{Shen}(1984)}]{Shen_nonlinearoptics}
\bibinfo{author}{\bibfnamefont{Y.~R.} \bibnamefont{Shen}},
  \emph{\bibinfo{title}{The Principles of Nonlinear Optics}}
  (\bibinfo{publisher}{Wiley}, \bibinfo{address}{New York},
  \bibinfo{year}{1984}), chaps. \bibinfo{chapter}{3, 7 and 9}.

\bibitem[{\citenamefont{Armstrong et~al.}(1962)\citenamefont{Armstrong,
  Bloembergen, Ducuing, and Pershan}}]{Armstrong62}
\bibinfo{author}{\bibfnamefont{J.~A.} \bibnamefont{Armstrong}},
  \bibinfo{author}{\bibfnamefont{N.}~\bibnamefont{Bloembergen}},
  \bibinfo{author}{\bibfnamefont{J.}~\bibnamefont{Ducuing}}, \bibnamefont{and}
  \bibinfo{author}{\bibfnamefont{P.~S.} \bibnamefont{Pershan}},
  \bibinfo{journal}{Phys. Rev.} \textbf{\bibinfo{volume}{127}},
  \bibinfo{pages}{1918} (\bibinfo{year}{1962}).

\bibitem[{\citenamefont{Franken and Ward}(1963)}]{Franken63}
\bibinfo{author}{\bibfnamefont{P.~A.} \bibnamefont{Franken}} \bibnamefont{and}
  \bibinfo{author}{\bibfnamefont{J.~F.} \bibnamefont{Ward}},
  \bibinfo{journal}{Rev. Mod. Phys.} \textbf{\bibinfo{volume}{35}},
  \bibinfo{pages}{23} (\bibinfo{year}{1963}).

\bibitem[{\citenamefont{Baldi et~al.}(1995)\citenamefont{Baldi, Aschieri, Nouh,
  Micheli, Ostrowsky, Delacourt, and Papuchon}}]{Baldi95}
\bibinfo{author}{\bibfnamefont{P.}~\bibnamefont{Baldi}},
  \bibinfo{author}{\bibfnamefont{P.}~\bibnamefont{Aschieri}},
  \bibinfo{author}{\bibfnamefont{S.}~\bibnamefont{Nouh}},
  \bibinfo{author}{\bibfnamefont{M.~D.} \bibnamefont{Micheli}},
  \bibinfo{author}{\bibfnamefont{D.~B.} \bibnamefont{Ostrowsky}},
  \bibinfo{author}{\bibfnamefont{D.}~\bibnamefont{Delacourt}},
  \bibnamefont{and} \bibinfo{author}{\bibfnamefont{M.}~\bibnamefont{Papuchon}},
  \bibinfo{journal}{IEEE J. Quantum Electron.} \textbf{\bibinfo{volume}{31}},
  \bibinfo{pages}{997} (\bibinfo{year}{1995}).

\bibitem[{\citenamefont{Myers et~al.}(1995)\citenamefont{Myers, Eckardt, Fejer,
  Byer, Bosenberg, and Pierce}}]{Myers95}
\bibinfo{author}{\bibfnamefont{L.~E.} \bibnamefont{Myers}},
  \bibinfo{author}{\bibfnamefont{R.~C.} \bibnamefont{Eckardt}},
  \bibinfo{author}{\bibfnamefont{M.~M.} \bibnamefont{Fejer}},
  \bibinfo{author}{\bibfnamefont{R.~L.} \bibnamefont{Byer}},
  \bibinfo{author}{\bibfnamefont{W.~R.} \bibnamefont{Bosenberg}},
  \bibnamefont{and} \bibinfo{author}{\bibfnamefont{J.~W.}
  \bibnamefont{Pierce}}, \bibinfo{journal}{J. Opt. Soc. Am. B}
  \textbf{\bibinfo{volume}{12}}, \bibinfo{pages}{2102} (\bibinfo{year}{1995}).

\bibitem[{\citenamefont{Fejer et~al.}(1992)\citenamefont{Fejer, Magel, Jundt,
  and Byer}}]{Fejer92}
\bibinfo{author}{\bibfnamefont{M.~M.} \bibnamefont{Fejer}},
  \bibinfo{author}{\bibfnamefont{G.~A.} \bibnamefont{Magel}},
  \bibinfo{author}{\bibfnamefont{D.~H.} \bibnamefont{Jundt}}, \bibnamefont{and}
  \bibinfo{author}{\bibfnamefont{R.~L.} \bibnamefont{Byer}},
  \bibinfo{journal}{IEEE J. Quantum Electron.} \textbf{\bibinfo{volume}{28}},
  \bibinfo{pages}{2631} (\bibinfo{year}{1992}).

\bibitem[{\citenamefont{Tanzilli et~al.}(2001)\citenamefont{Tanzilli,
  Riedmatten, Tittel, Zbinden, Baldi, Micheli, Ostrowsky, and
  Gisin}}]{Tanzilli01_EL}
\bibinfo{author}{\bibfnamefont{S.}~\bibnamefont{Tanzilli}},
  \bibinfo{author}{\bibfnamefont{H.~D.} \bibnamefont{Riedmatten}},
  \bibinfo{author}{\bibfnamefont{W.}~\bibnamefont{Tittel}},
  \bibinfo{author}{\bibfnamefont{H.}~\bibnamefont{Zbinden}},
  \bibinfo{author}{\bibfnamefont{P.}~\bibnamefont{Baldi}},
  \bibinfo{author}{\bibfnamefont{M.~D.} \bibnamefont{Micheli}},
  \bibinfo{author}{\bibfnamefont{D.~B.} \bibnamefont{Ostrowsky}},
  \bibnamefont{and} \bibinfo{author}{\bibfnamefont{N.}~\bibnamefont{Gisin}},
  \bibinfo{journal}{Electron. Lett.} \textbf{\bibinfo{volume}{37}},
  \bibinfo{pages}{26} (\bibinfo{year}{2001}).

\bibitem[{\citenamefont{Pe{\v r}ina et~al.}(1994)\citenamefont{Pe{\v r}ina,
  Hradil, and Jur{\v c}o}}]{Perina94}
\bibinfo{author}{\bibfnamefont{J.}~\bibnamefont{Pe{\v r}ina}},
  \bibinfo{author}{\bibfnamefont{Z.}~\bibnamefont{Hradil}}, \bibnamefont{and}
  \bibinfo{author}{\bibfnamefont{B.}~\bibnamefont{Jur{\v c}o}},
  \emph{\bibinfo{title}{Quantum Optics and Fundamentals of Physics}}
  (\bibinfo{publisher}{Kluwer}, \bibinfo{address}{Boston},
  \bibinfo{year}{1994}), chaps. \bibinfo{chapter}{7 and 8}.

\bibitem[{\citenamefont{Mandel and Wolf}(1995)}]{Mandel95}
\bibinfo{author}{\bibfnamefont{L.}~\bibnamefont{Mandel}} \bibnamefont{and}
  \bibinfo{author}{\bibfnamefont{E.}~\bibnamefont{Wolf}},
  \emph{\bibinfo{title}{Optical Coherence and Quantum Optics}}
  (\bibinfo{publisher}{Cambridge Univ. Press}, \bibinfo{address}{Cambridge},
  \bibinfo{year}{1995}), chap.~\bibinfo{chapter}{22}.

\bibitem[{\citenamefont{Banaszek et~al.}(2001)\citenamefont{Banaszek, U'Ren,
  and Walmsley}}]{Banaszek01}
\bibinfo{author}{\bibfnamefont{K.}~\bibnamefont{Banaszek}},
  \bibinfo{author}{\bibfnamefont{A.~B.} \bibnamefont{U'Ren}}, \bibnamefont{and}
  \bibinfo{author}{\bibfnamefont{I.~A.} \bibnamefont{Walmsley}},
  \bibinfo{journal}{Opt. Lett.} \textbf{\bibinfo{volume}{26}},
  \bibinfo{pages}{1367} (\bibinfo{year}{2001}).

\bibitem[{\citenamefont{Rossi and Berger}(2002)}]{DeRossi02}
\bibinfo{author}{\bibfnamefont{A.~D.} \bibnamefont{Rossi}} \bibnamefont{and}
  \bibinfo{author}{\bibfnamefont{V.}~\bibnamefont{Berger}},
  \bibinfo{journal}{Phys. Rev. Lett.} \textbf{\bibinfo{volume}{88}},
  \bibinfo{pages}{043901} (\bibinfo{year}{2002}).

\bibitem[{\citenamefont{Atat{\" u}re et~al.}(2001)\citenamefont{Atat{\" u}re,
  {Di Giuseppe}, Shaw, Saleh, Sergienko, and Teich}}]{Atature01_PRA}
\bibinfo{author}{\bibfnamefont{M.}~\bibnamefont{Atat{\" u}re}},
  \bibinfo{author}{\bibfnamefont{G.}~\bibnamefont{{Di Giuseppe}}},
  \bibinfo{author}{\bibfnamefont{M.~D.} \bibnamefont{Shaw}},
  \bibinfo{author}{\bibfnamefont{B.~E.~A.} \bibnamefont{Saleh}},
  \bibinfo{author}{\bibfnamefont{A.~V.} \bibnamefont{Sergienko}},
  \bibnamefont{and} \bibinfo{author}{\bibfnamefont{M.~C.} \bibnamefont{Teich}},
  \bibinfo{journal}{quant-ph/0111024}  (\bibinfo{year}{2001}),
  \bibinfo{note}{{P}hys. {R}ev. {A}, submitted}.

\bibitem[{\citenamefont{Klyshko}(1980)}]{Klyshko80}
\bibinfo{author}{\bibfnamefont{D.~N.} \bibnamefont{Klyshko}},
  \emph{\bibinfo{title}{Photons and Nonlinear Optics}}
  (\bibinfo{publisher}{Nauka}, \bibinfo{address}{Moscow},
  \bibinfo{year}{1980}), chaps. \bibinfo{chapter}{1 and 6} 
  \bibinfo{note}{[Translation: Gordon and Breach, 1988]}.

\bibitem[{\citenamefont{{Di Giuseppe} et~al.}(2002)\citenamefont{{Di Giuseppe},
  Atat{\" u}re, Shaw, Sergienko, Saleh, and Teich}}]{Giuseppe02}
\bibinfo{author}{\bibfnamefont{G.}~\bibnamefont{{Di Giuseppe}}},
  \bibinfo{author}{\bibfnamefont{M.}~\bibnamefont{Atat{\" u}re}},
  \bibinfo{author}{\bibfnamefont{M.~D.} \bibnamefont{Shaw}},
  \bibinfo{author}{\bibfnamefont{A.~V.} \bibnamefont{Sergienko}},
  \bibinfo{author}{\bibfnamefont{B.~E.~A.} \bibnamefont{Saleh}},
  \bibnamefont{and} \bibinfo{author}{\bibfnamefont{M.~C.} \bibnamefont{Teich}},
  \bibinfo{journal}{quant-ph/0112140}  (\bibinfo{year}{2002}),
  \bibinfo{note}{{P}hys. {R}ev. {A}, submitted}.

\bibitem[{\citenamefont{Saleh and Teich}(1991)}]{Saleh91_waveguides}
\bibinfo{author}{\bibfnamefont{B.~E.~A.} \bibnamefont{Saleh}} \bibnamefont{and}
  \bibinfo{author}{\bibfnamefont{M.~C.} \bibnamefont{Teich}},
  \emph{\bibinfo{title}{Fundamentals of Photonics}}
  (\bibinfo{publisher}{Wiley}, \bibinfo{address}{New York},
  \bibinfo{year}{1991}), chap.~\bibinfo{chapter}{7}.

\bibitem[{\citenamefont{Dmitriev et~al.}(1995)\citenamefont{Dmitriev,
  Gurzadyan, and Nikogosyan}}]{Crystals.Handbook}
\bibinfo{author}{\bibfnamefont{V.~G.} \bibnamefont{Dmitriev}},
  \bibinfo{author}{\bibfnamefont{G.~G.} \bibnamefont{Gurzadyan}},
  \bibnamefont{and} \bibinfo{author}{\bibfnamefont{D.~N.}
  \bibnamefont{Nikogosyan}}, \emph{\bibinfo{title}{The Handbook of Nonlinear
  Optical Crystals}}, \bibinfo{edition}{2nd} ed. (\bibinfo{publisher}{Springer-Verlag},
  \bibinfo{address}{New York}, \bibinfo{year}{1995}), vol.~\bibinfo{volume}{64}
  of \emph{\bibinfo{series}{Springer Series in Optical Sciences}},
  chap.~\bibinfo{chapter}{3}.

\bibitem[{\citenamefont{Batchko et~al.}(1999)\citenamefont{Batchko, Fejer,
  Byer, Woll, Wallenstein, Shur, and Erman}}]{Batchko99}
\bibinfo{author}{\bibfnamefont{R.~G.} \bibnamefont{Batchko}},
  \bibinfo{author}{\bibfnamefont{M.~M.} \bibnamefont{Fejer}},
  \bibinfo{author}{\bibfnamefont{R.~L.} \bibnamefont{Byer}},
  \bibinfo{author}{\bibfnamefont{D.}~\bibnamefont{Woll}},
  \bibinfo{author}{\bibfnamefont{R.}~\bibnamefont{Wallenstein}},
  \bibinfo{author}{\bibfnamefont{V.~Y.} \bibnamefont{Shur}}, \bibnamefont{and}
  \bibinfo{author}{\bibfnamefont{L.}~\bibnamefont{Erman}},
  \bibinfo{journal}{Opt. Lett.} \textbf{\bibinfo{volume}{24}},
  \bibinfo{pages}{1293} (\bibinfo{year}{1999}).

\bibitem[{\citenamefont{Ding et~al.}(1995)\citenamefont{Ding, Lee, and
  Khurgin}}]{Ding95}
\bibinfo{author}{\bibfnamefont{Y.~J.} \bibnamefont{Ding}},
  \bibinfo{author}{\bibfnamefont{S.~J.} \bibnamefont{Lee}}, \bibnamefont{and}
  \bibinfo{author}{\bibfnamefont{J.~B.} \bibnamefont{Khurgin}},
  \bibinfo{journal}{Phys. Rev. Lett.} \textbf{\bibinfo{volume}{75}},
  \bibinfo{pages}{429} (\bibinfo{year}{1995}).

\bibitem[{\citenamefont{Ghatak and Thyagarajan}(1998)}]{Ghatak98}
\bibinfo{author}{\bibfnamefont{A.}~\bibnamefont{Ghatak}} \bibnamefont{and}
  \bibinfo{author}{\bibfnamefont{K.}~\bibnamefont{Thyagarajan}},
  \emph{\bibinfo{title}{Introduction to Fiber Optics}}
  (\bibinfo{publisher}{Cambridge}, \bibinfo{address}{New York},
  \bibinfo{year}{1998}), chap.~\bibinfo{chapter}{7}.

\bibitem[{\citenamefont{Bonfrate et~al.}(1999)\citenamefont{Bonfrate, Pruneri,
  Kazansky, Tapster, and Rarity}}]{Bonfrate99}
\bibinfo{author}{\bibfnamefont{G.}~\bibnamefont{Bonfrate}},
  \bibinfo{author}{\bibfnamefont{V.}~\bibnamefont{Pruneri}},
  \bibinfo{author}{\bibfnamefont{P.~G.} \bibnamefont{Kazansky}},
  \bibinfo{author}{\bibfnamefont{P.}~\bibnamefont{Tapster}}, \bibnamefont{and}
  \bibinfo{author}{\bibfnamefont{J.~G.} \bibnamefont{Rarity}},
  \bibinfo{journal}{Appl. Phys. Lett.} \textbf{\bibinfo{volume}{16}},
  \bibinfo{pages}{2356} (\bibinfo{year}{1999}).

\bibitem[{\citenamefont{Kazansky and Pruneri}(1997)}]{Kazansky97}
\bibinfo{author}{\bibfnamefont{P.~G.} \bibnamefont{Kazansky}} \bibnamefont{and}
  \bibinfo{author}{\bibfnamefont{V.}~\bibnamefont{Pruneri}},
  \bibinfo{journal}{J. Opt. Soc. Am. B} \textbf{\bibinfo{volume}{14}},
  \bibinfo{pages}{3170} (\bibinfo{year}{1997}).

\bibitem[{\citenamefont{Wang et~al.}(1999)\citenamefont{Wang, Pasiskevivius,
  Hellstr{\" o}m, Laurell, and Karlsson}}]{Wang99}
\bibinfo{author}{\bibfnamefont{S.}~\bibnamefont{Wang}},
  \bibinfo{author}{\bibfnamefont{V.}~\bibnamefont{Pasiskevivius}},
  \bibinfo{author}{\bibfnamefont{J.}~\bibnamefont{Hellstr{\" o}m}},
  \bibinfo{author}{\bibfnamefont{F.}~\bibnamefont{Laurell}}, \bibnamefont{and}
  \bibinfo{author}{\bibfnamefont{H.}~\bibnamefont{Karlsson}},
  \bibinfo{journal}{Opt. Lett.} \textbf{\bibinfo{volume}{24}},
  \bibinfo{pages}{978} (\bibinfo{year}{1999}).

\end{thebibliography}

%%%%%%%%%%%%%%%%%%%%%%%%%%%%%%%%%
%       FIGURE CAPTIONS         %
%%%%%%%%%%%%%%%%%%%%%%%%%%%%%%%%%
\clearpage
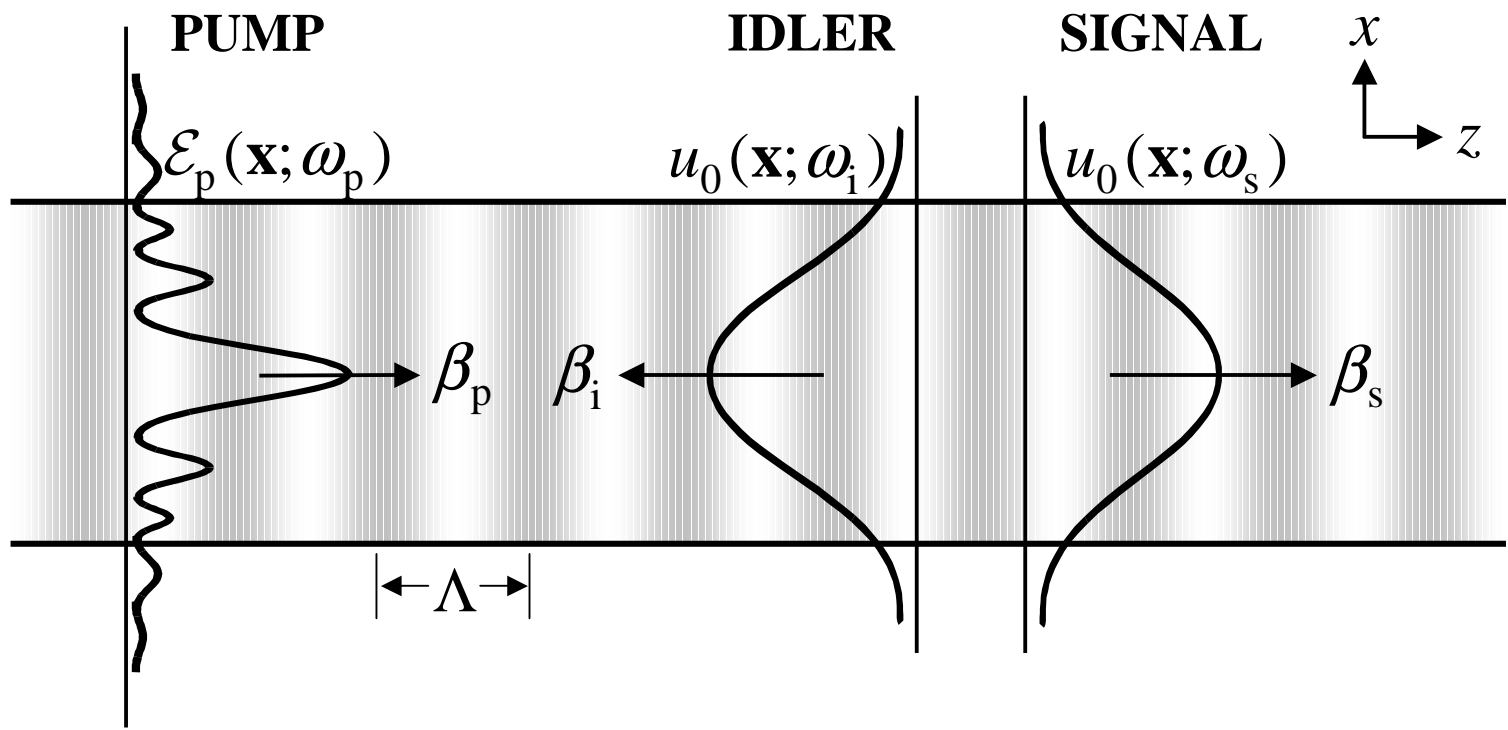
\begin{figure}
\caption{Schematic of a slab waveguide structure with modulated
nonlinearity (illustrated as striations). The pump has an
arbitrary spatial profile whereas the signal and idler beams are
assumed to occupy the fundamental TE electric-field modes of the
waveguide.} \label{fig:schematic}
\end{figure}

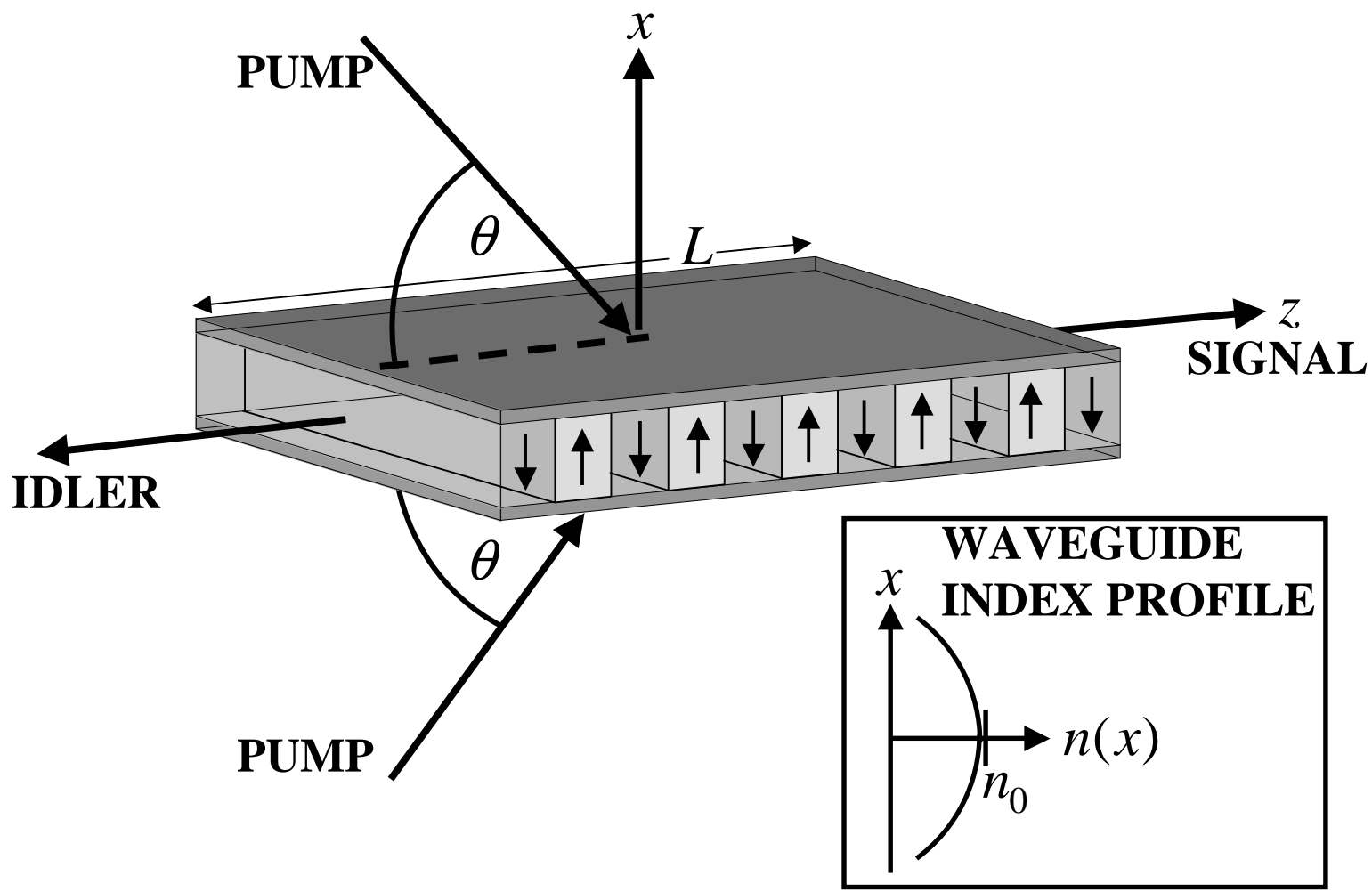
\begin{figure}
\caption{Schematic of a PPLN slab waveguide with the plane-wave
pump beams incident at the angles of incidence $\pm \theta$. The
pump is unguided whereas the counter-propagating signal and idler
beams are assumed to populate the fundamental mode of the
waveguide. The medium comprising the waveguide is taken to have a
parabolic index profile $n(x)$ as shown in the inset and to have a
nonlinear susceptibility $\chi^{(2)}(z)$ that has a single period.
The illustration shows a square-wave modulation of $\chi^{(2)}$;
the arrows indicate regions of positive and negative values of the
nonlinear coefficient.} \label{fig:unguided-pump}
\end{figure}

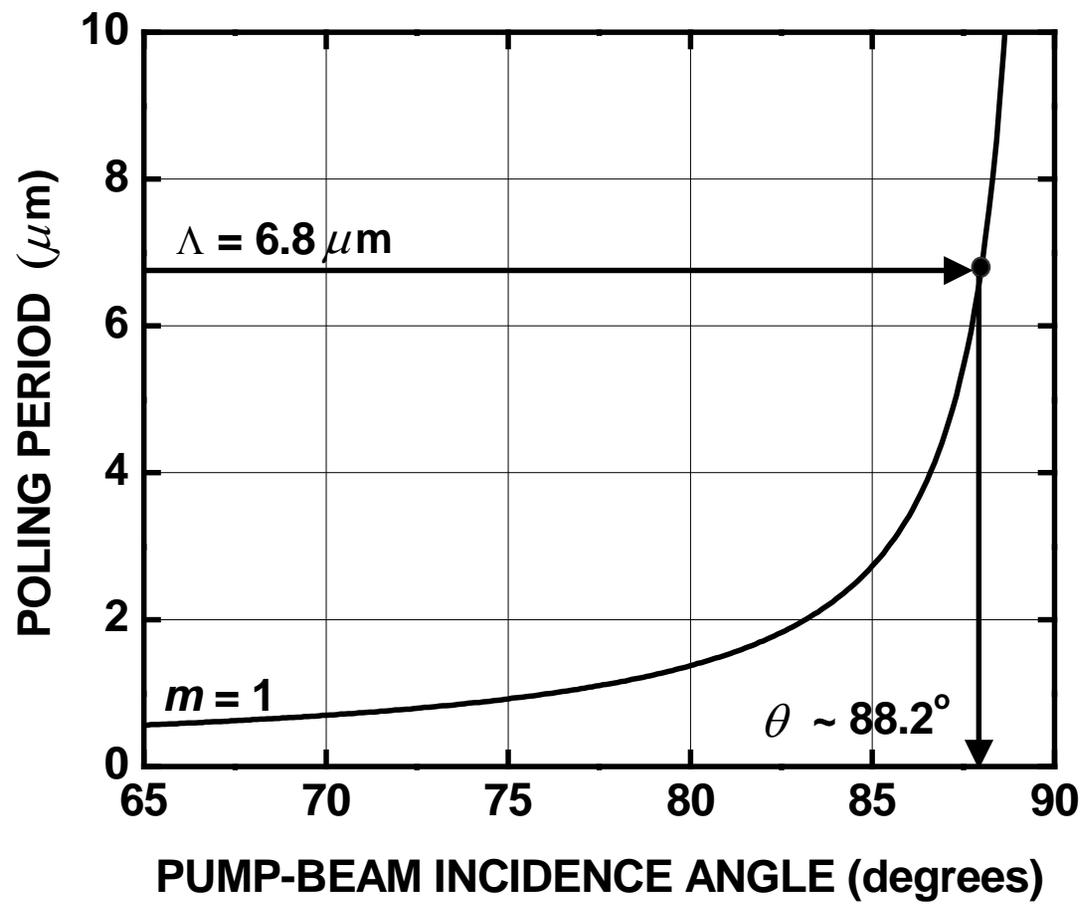
\begin{figure}
\caption{Degenerate tuning curve for perfect QPM ($\Delta \beta' =
0$): Poling period $\Lambda$ ($m$ = 1) vs. pump-beam incidence
angle $\theta$ for degenerate down-conversion in a PPLN waveguide
with counter-propagating signal and idler photons at $\lambda_{\rm
s} = \lambda_{\rm i} =$ 1064 nm. The pump wavelength $\lambda_{\rm
p} =$ 532~nm.} \label{fig:degen}
\end{figure}

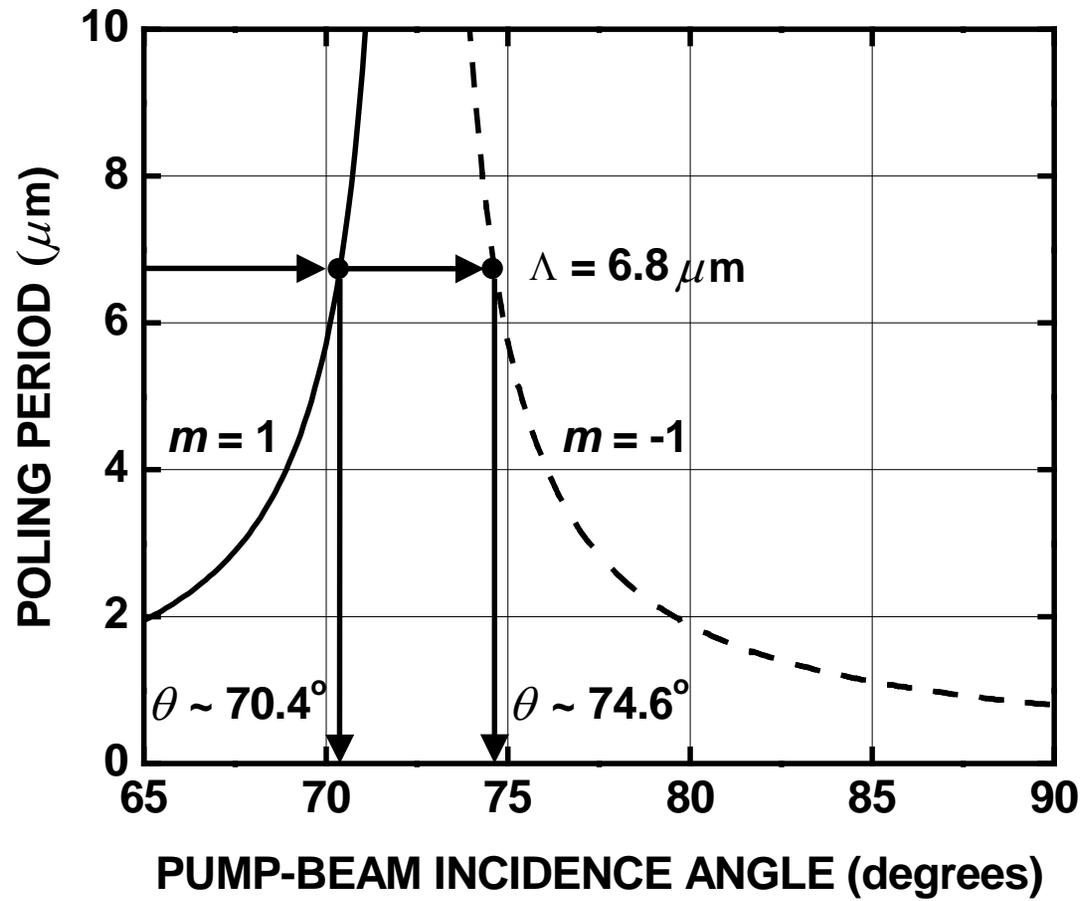
\begin{figure}
\caption{Non-degenerate tuning curve for perfect QPM ($\Delta
\beta' = 0$): Poling period $\Lambda$ ($m$ = $\pm 1$) vs.
pump-beam incidence angle $\theta$ for non-degenerate
down-conversion in a PPLN waveguide with counter-propagating
signal and idler photons at $\lambda_{\rm s} =$ 810 nm and
$\lambda_{\rm i} =$ 1550 nm. The pump wavelength $\lambda_{\rm p}
=$ 532~nm.} \label{fig:nondegen}
\end{figure}

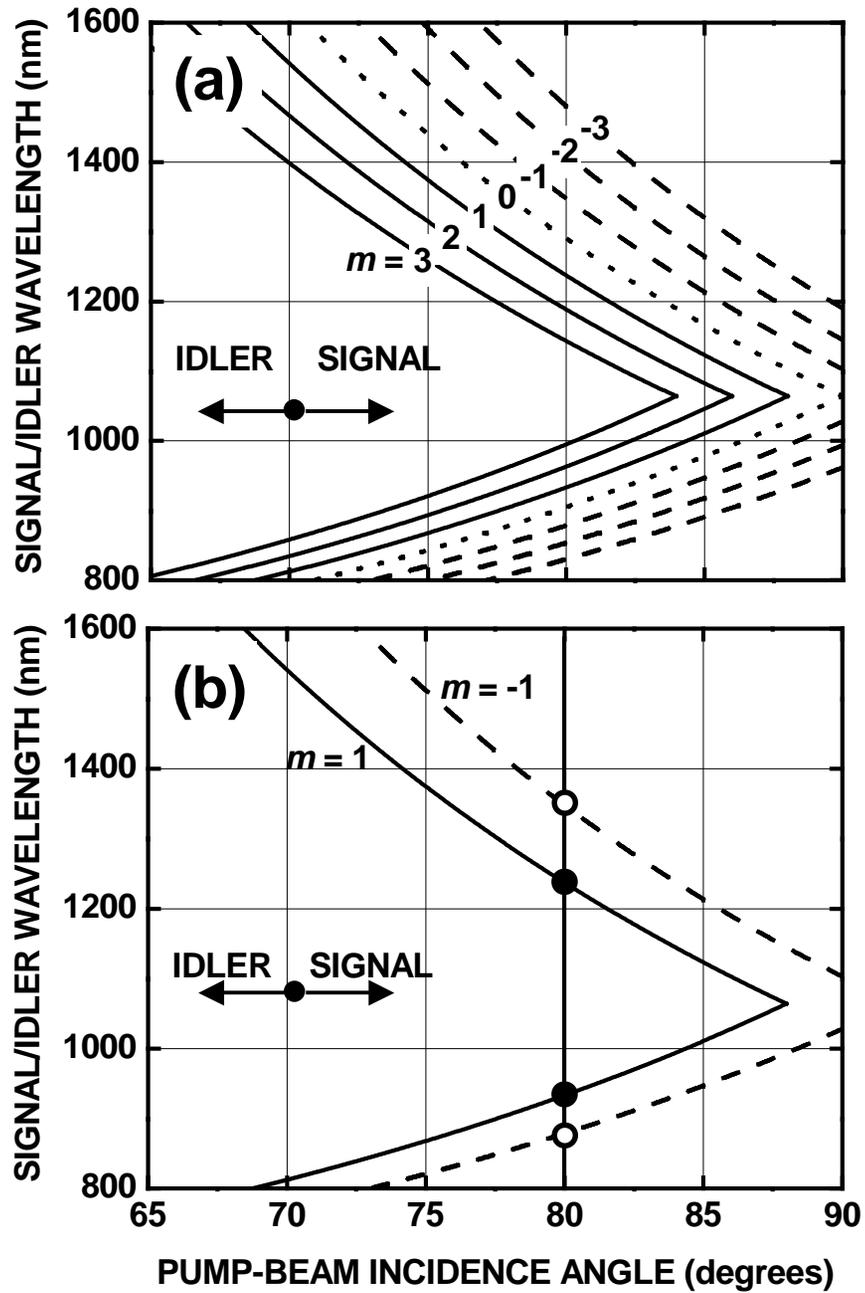
\begin{figure}
\caption{Tuning curves for perfect QPM ($\Delta \beta' = 0$) with
various values of $m$:  (a) Signal and idler wavelengths vs.
pump-beam incidence angle $\theta$ for grating vector orders $m =$
0, $\pm$1, $\pm$2, and $\pm$3 in a PPLN waveguide with a poling
period $\Lambda = 6.8~\mu\rm{m}$. (b) Subset of tuning curves in
(a) for $m=$~$\pm$1 with the poling period $\Lambda =
6.8~\mu\rm{m}$. The signal photon propagates in the positive
$z$-direction and the idler photon counter-propagates in the
negative $z$-direction as shown by the inset. The pump wavelength
$\lambda_{\rm p} =$ 532~nm in both (a) and (b). Positive and
negative values of $m$ are indicated by solid and dashed curves,
respectively. The case where $m=$~0 is indicated by a dotted
curve. Circles in (b) indicate the two signal-idler combinations
possible for a pump-beam incidence angle of $80^\circ$: 880
nm/1350 nm ($m=$~-1, dashed curve, open circles) and 930 nm/1240
nm ($m=$~+1, solid curve, solid circles). Since these plots only
consider the case of perfect QPM, they lack the spectral
information of the signal and idler photons, which will be
presented in Fig.~\ref{fig:2D_counter_SPDC_spec}.}
\label{fig:lambda_v_angle}
\end{figure}

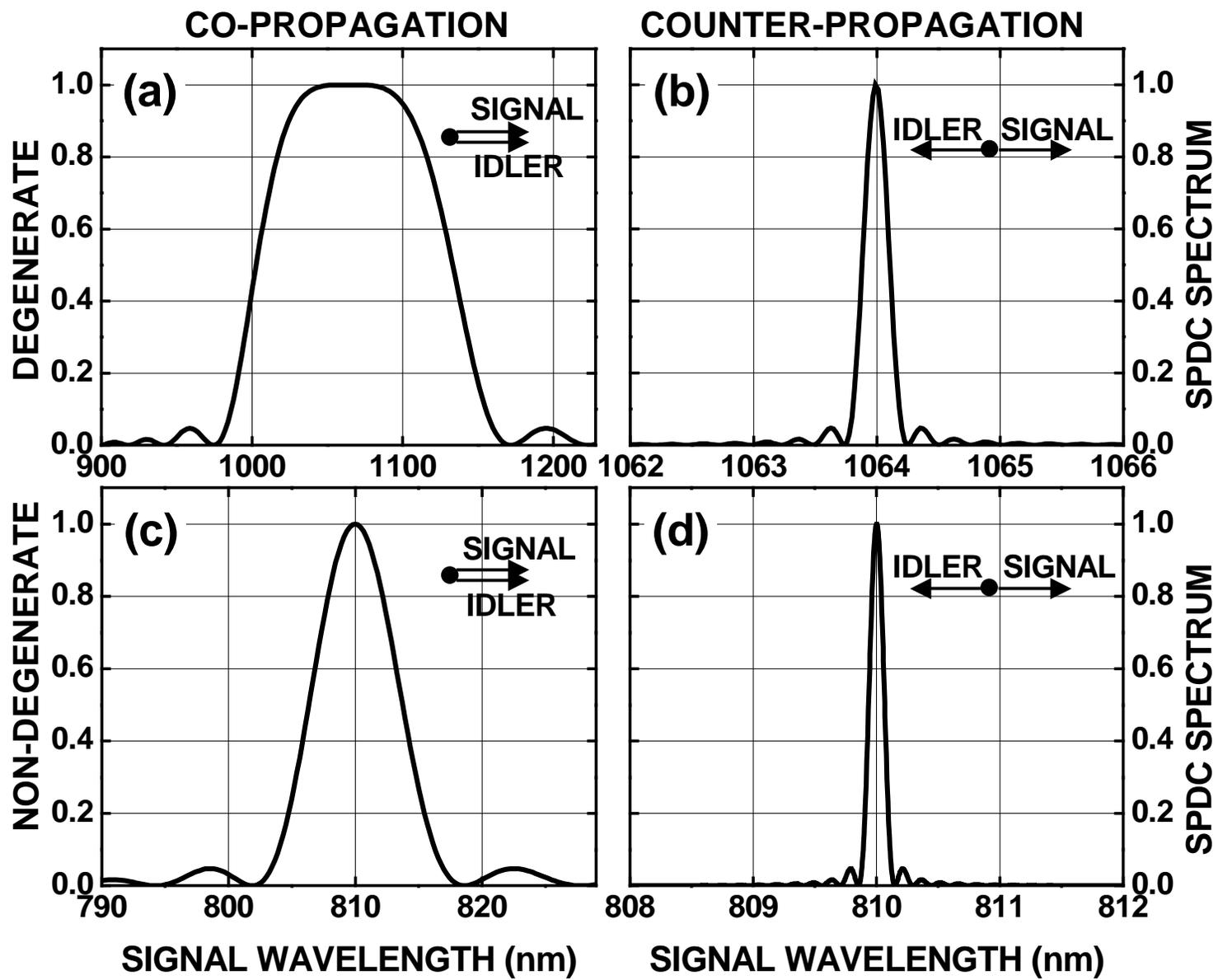
\begin{figure}
\caption{SPDC spectrum vs. signal wavelength $\lambda_{\rm s}$ for
phase-matching conditions that allow (a) degenerate,
co-propagating ($\theta = 0^\circ$); (b) degenerate,
counter-propagating ($\theta = 88.2^\circ$); (c) non-degenerate,
co-propagating ($\Lambda = 7.4~\mu{\rm m}$, $\theta = 0^\circ$);
and (d) non-degenerate, counter-propagating ($\theta =
69.7^\circ$) signal and idler photons in a PPLN waveguide of 1-mm
length.  The directions of propagation for the signal and idler
photons in the various cases is shown by the insets. Note the
different abscissa scales for the co-propagating and
counter-propagating panels.  The full-width-half-maximum (FWHM)
for the co-propagating case is 130~nm for degenerate photons
versus 7.3~nm for a non-degenerate signal photon at 810~nm.  In
the counter-propagating case, the FWHM for degenerate photons is
0.23~nm versus 0.13~nm for a non-degenerate signal photon at
810~nm. The poling period in (a), (b), and (d) is $\Lambda =
6.8~\mu\rm{m}$, whereas for (c) it is $\Lambda = 7.4~\mu\rm{m}$ to
achieve a non-degenerate solution for a $0^\circ$ angle of
incidence. The pump wavelength $\lambda_{\rm p} = 532$~nm and the
non-degenerate idler photon $\lambda_{\rm i} = 1550$~nm.}
\label{fig:SPDC_degen}
\end{figure}

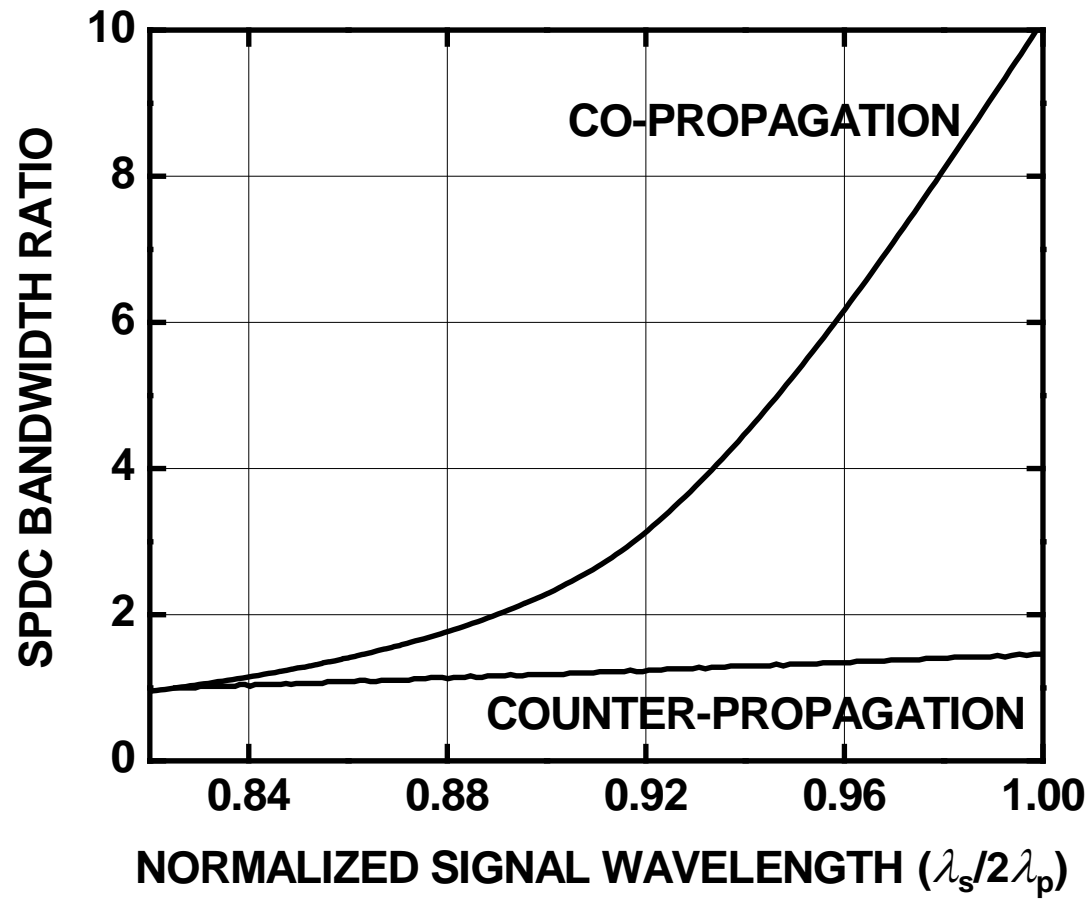
\begin{figure}
\caption{SPDC bandwidth ratio vs. normalized signal wavelength.
The bandwidth ratio is obtained by dividing by the bandwidth at
$\lambda_{\rm s}$ = 880 nm (0.83 on the abscissa), which is the
shortest wavelength used for this simulation. The signal
wavelength is normalized to the degenerate wavelength
$2\lambda_{\rm p}=$ 1064 nm. The pump wavelength $\lambda_{\rm p}
= 532$~nm. This plot illustrates the relative increase in
bandwidth for co-propagating and counter-propagating photons as
the nonlinear interaction approaches degeneracy at $\lambda_{\rm
s}/2\lambda_{\rm p}=$~1.00. } \label{fig:SPDC_nondegen}
\end{figure}

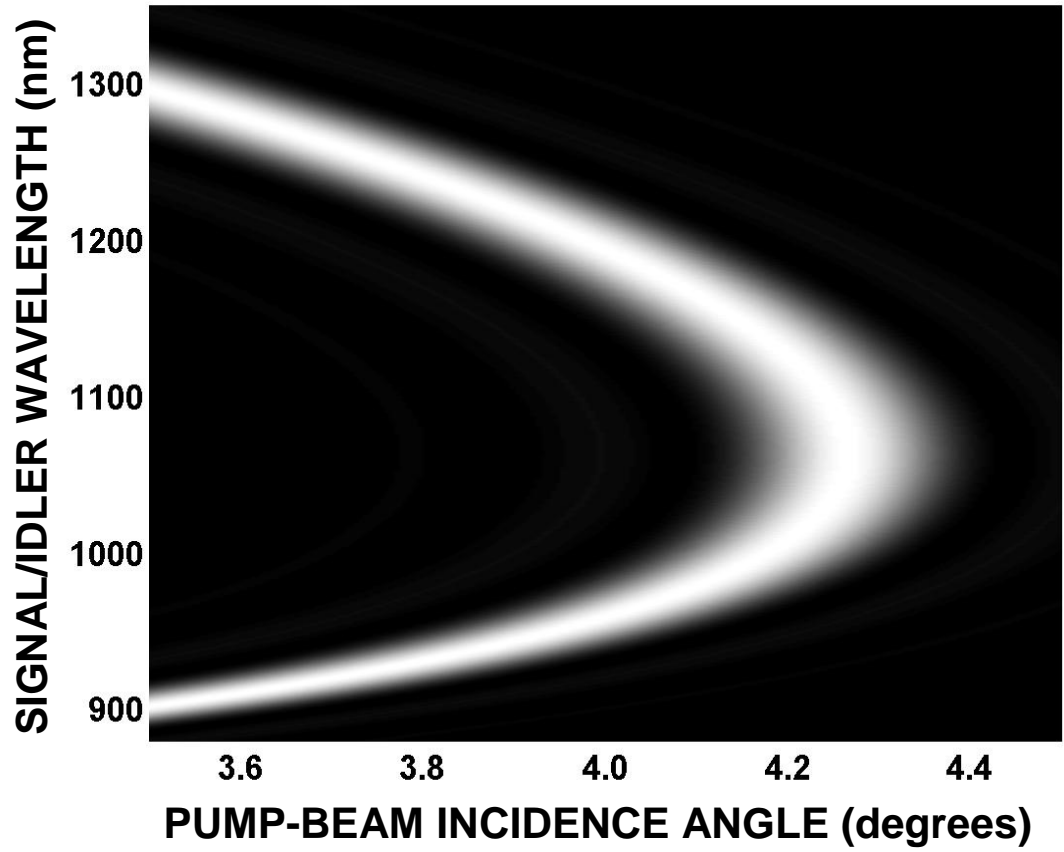
\begin{figure}
\caption{Spectrum of signal and idler vs. pump-beam incidence
angle for co-propagating beams in a 1-mm PPLN waveguide with a
poling period $\Lambda = 7.4~\mu\rm{m}$ and positive grating
vector $K_m$ ($m=1$). The pump wavelength $\lambda_{\rm p} =
532$~nm.} \label{fig:2D_SPDC_spec}
\end{figure}

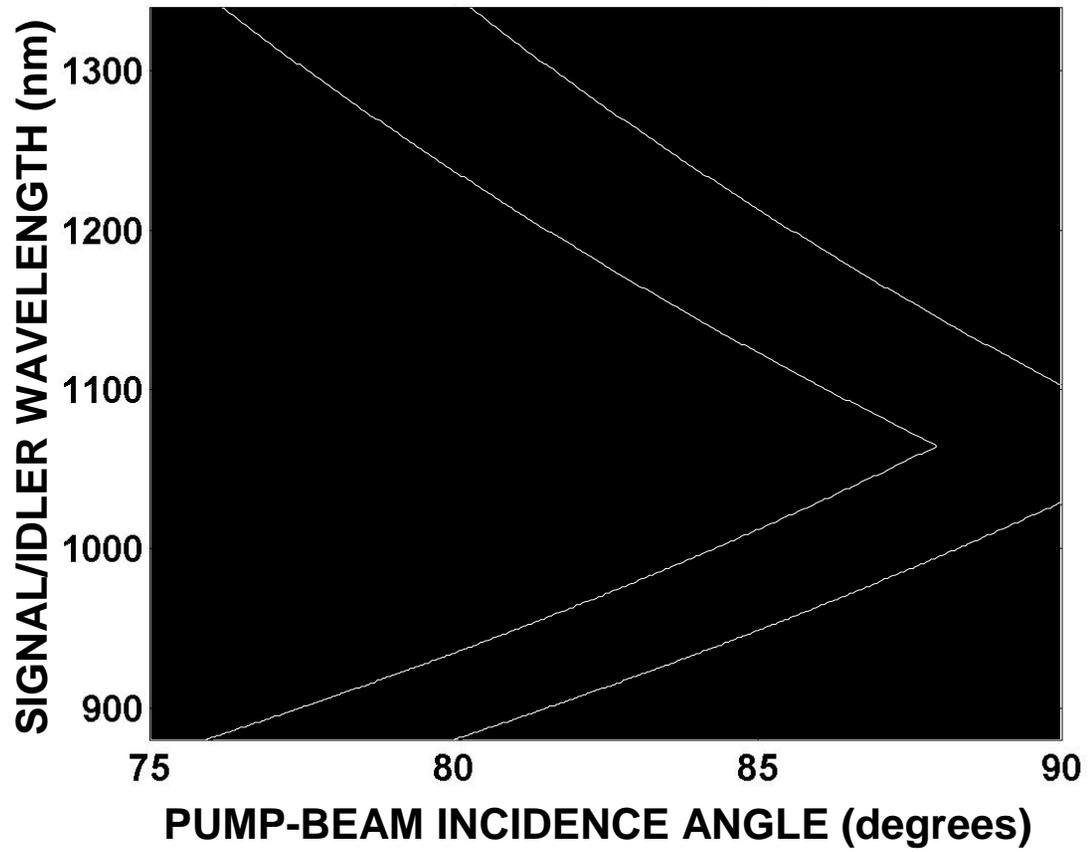
\begin{figure}
\caption{Spectrum of signal and idler vs. pump-beam incidence
angle for counter-propagating beams in a 1-mm PPLN waveguide with
a poling period $\Lambda = 6.8~\mu\rm{m}$ and $m=\pm$1. The pump
wavelength $\lambda_{\rm p} = 532$~nm.  The narrow bandwidth of
the down-converted light is evident for all signal and idler
combinations and does not change appreciably as the photons
approach degeneracy, unlike the results shown in
Fig.~\ref{fig:2D_SPDC_spec} for co-propagating beams.  Although
this plot includes all the information pertaining to the spectrum
of the down-converted light, it is visually indistinguishable from
the plot presented in Fig.~\ref{fig:lambda_v_angle}(b).}
\label{fig:2D_counter_SPDC_spec}
\end{figure}

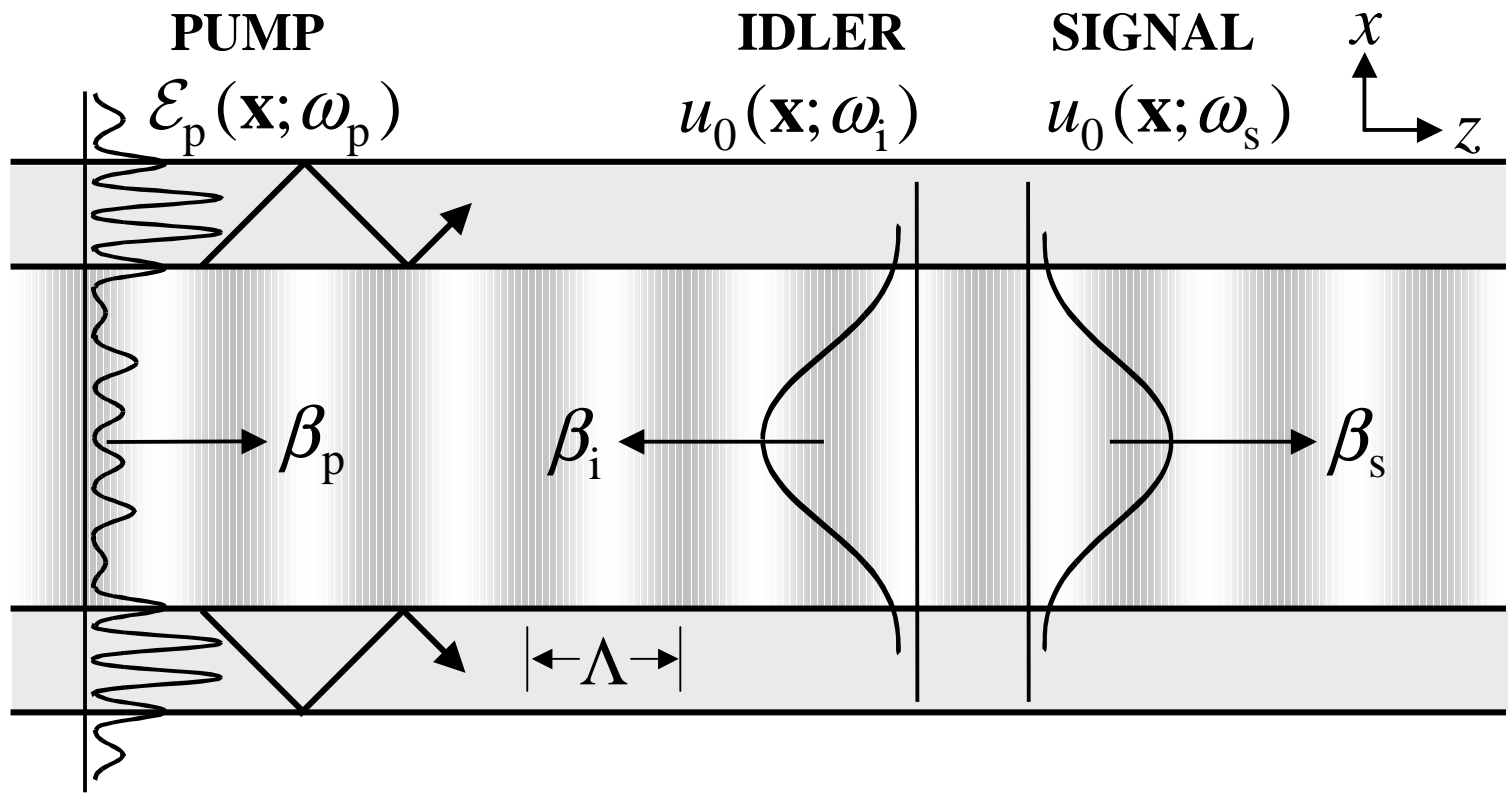
\begin{figure}
\caption{Meridional cross-section of an optical fiber with a
periodically modulated nonlinear medium $\chi^{(2)}$ that
comprises the core. The pump beam is guided within the fiber
cladding. By coupling to the core, spontaneous parametric
down-conversion yields counter-propagating signal and idler
photons.} \label{fig:nonlinear-fiber}
\end{figure}

\end{document}